\documentclass[sigconf]{acmart}

\usepackage{amsmath}
\usepackage{amsfonts}
\usepackage{bm}
\usepackage{color}
\usepackage{graphicx}
\usepackage{url}
\usepackage{setspace}
\usepackage{subcaption}
\usepackage{booktabs}
\usepackage{mathtools}
\usepackage{multirow}
\usepackage{float}
\usepackage{caption} 
\usepackage{algorithm}
\usepackage{algorithmic}
\usepackage{cleveref}
\usepackage{makecell}
\usepackage{amsthm}
\usepackage{amsmath}
\usepackage{amsfonts}
\usepackage{xcolor,colortbl}
\usepackage{tabularx}
\definecolor{lightlightgray}{gray}{0.85}
\definecolor{lllgray}{gray}{0.9}
\definecolor{llllgray}{gray}{0.95}
\crefname{equation}{Eq.}{Eq.}
\crefname{section}{Section}{Sections}
\crefname{subsection}{Section}{Sections}
\crefname{subsubsection}{Section}{Sections}
\crefname{figure}{Figure}{Figures}
\crefname{table}{Table}{Tables}
\crefname{subfigure}{Figure}{Figures}
\crefname{algocf}{Algorithm}{Algorithms}

\newtheorem*{hypothesis}{Hypothesis}

\usepackage{enumitem}
\setlist{after=\vspace{0\baselineskip},leftmargin=12pt}
\copyrightyear{2023}
\acmYear{2023}
\setcopyright{acmlicensed}\acmConference[WWW '23]{Proceedings of the ACM Web Conference 2023}{May 1--5, 2023}{Austin, TX, USA}
\acmBooktitle{Proceedings of the ACM Web Conference 2023 (WWW '23), May 1--5, 2023, Austin, TX, USA}
\acmPrice{15.00}
\acmDOI{10.1145/3543507.3583400}
\acmISBN{978-1-4503-9416-1/23/04}

\begin{document}

\title{Large-Scale Analysis of New Employee Network Dynamics}

\author{Yulin Yu}
\affiliation{
  \institution{University of Michigan}
  \country{Ann Arbor, MI, USA}
  }
\email{yulinyu@umich.edu}

\author{Longqi Yang}
\affiliation{
  \institution{Microsoft}
  \country{Redmond, WA, USA}
  }
\email{longqi.yang@microsoft.com}

\author{Si\^an Lindley}
\affiliation{
  \institution{Microsoft}
  \country{Cambridge, United Kingdom}
  }
\email{sianl@microsoft.com}

\author{Mengting Wan}
\affiliation{
  \institution{Microsoft}
  \country{Redmond, WA, USA}
  }
\email{mengting.wan@microsoft.com}

\renewcommand{\shortauthors}{Yu, et al.}

\begin{abstract}
The COVID-19 pandemic has accelerated digital transformations across industries, but also introduced new challenges into workplaces, including the difficulties of effectively socializing with colleagues when working remotely. This challenge is exacerbated for new employees who need to develop workplace networks from the outset. 
In this paper,  by analyzing a large-scale telemetry dataset of more than 10,000 Microsoft employees who joined the company in the first three months of 2022,
we describe how new employees interact and telecommute with their colleagues during their ``onboarding'' period. 
Our results reveal that although new hires are gradually expanding networks over time, there still exists significant gaps between their network statistics and those of tenured employees even after the six-month onboarding phase. We also observe that heterogeneity exists among new employees in how their networks change over time, where employees whose job tasks do not necessarily require extensive and diverse connections could be at a disadvantaged position in this onboarding process. By investigating how web-based people recommendations in organizational knowledge base facilitate new employees naturally expand their networks, we also demonstrate the potential of web-based applications for addressing the aforementioned socialization challenges.
Altogether, our findings provide insights on new employee network dynamics in remote and hybrid work environments, which may help guide organizational leaders and web application developers on quantifying and improving the socialization experiences of new employees in digital workplaces.

\end{abstract}

\begin{CCSXML}
<ccs2012>
   <concept>
       <concept_id>10010405.10010455.10010461</concept_id>
       <concept_desc>Applied computing~Sociology</concept_desc>
       <concept_significance>500</concept_significance>
       </concept>
   <concept>
       <concept_id>10002951.10003227.10003351</concept_id>
       <concept_desc>Information systems~Data mining</concept_desc>
       <concept_significance>300</concept_significance>
       </concept>
 </ccs2012>
\end{CCSXML}

\ccsdesc[500]{Applied computing}
\ccsdesc[300]{Information systems~Data mining}



\keywords{Future of Work, Onboarding, Organizations, Social Networks,  Temporal Dynamics}


\maketitle

\section{Introduction}

New employee onboarding often refers to the process where new hires are integrated into the organization. This process is widely acknowledged to be vital for organizations and positively related to employee satisfaction, productivity and retention \cite{cable1996person,cable2013breaking,caldwell2018new,ellis2014new}. 
A successful new employee onboarding process not only involves knowledge acquisition, skills training, function understanding, but also encourages new hires to adapt to the organizational culture and socialize with their authentic selves, therefore establishing the sense of belonging \cite{cable1996person}. Although there exist extensive studies on conceptualizing this organizational socialization process \cite{ellis2014new,katz1997human,hall1996careers}, empirical evidence about how new hires communicate and socialize with other employees throughout their onboarding period is very limited to date.

The COVID-19 pandemic has caused a fundamental shift to remote work for many information workers \cite{barrero2021working}. It has accelerated the digital transformation of many workplaces, where employees became reliant on telecommunications such as remote meeting and instant messaging. Despite the flexibility introduced by this working paradigm change, it also brings new onboarding challenges, largely due to the lack of spontaneous and informal in-person conversations \cite{butler2021challenges, michel2021flattening, brucks2022virtual}. Studies reveal that the collaboration network of information workers became more static and siloed with the shift to firm-wide remote work, which makes it more difficult for employees to acquire and share novel information across the network \cite{Yang2021-rr}. This struggle for organizational socialization can be aggravated on new employees who need to not only acquire necessary knowledge and skills in a short time, but also develop their workplace networks from scratch. For example, recent research on software developers indicates that new hires struggled to ask for help while working remotely (due to the isolation between teammates, scheduling difficulties, the lack of hallway conversations, etc.) and communicated with fewer people overall compared to those who joined pre-pandemic \cite{Rodeghero2021-rr, Teevan2020-ea}.

This implies the need for different tools and approaches to supporting onboarding in remote settings. Additionally, the shift to remote work  enables new opportunities to explore the impact of such tools, by  empirically quantifying the socialization behavior of new employees from their workplace digital traces, and the extent to which they can be augmented or improved by web applications.
In this paper, we empirically study the socialization behavior of new employees on a large-scale post-pandemic workplace communication dataset from Microsoft, which includes rich telemetry data of employees on their remote video/audio meetings, emails, and instant messages. In addition, we measure the usage patterns of a web-based people recommendation engine in the workplace and demonstrate its potential in addressing onboarding challenges. The network characteristics we measure include:
\begin{itemize}
\item[1)] The number of distinct connections (or social ties) each employee maintains in each week; 
\item[2)] The frequency in which each individual communicates with other employees every week;
\item[3)] The extent to which an employee bridges communities within an organization.
\end{itemize}
The first two dimensions approximate the number of potential information sources and the total amount of information exchanged with the focal employee. The third dimension captures one's network structural diversity, which has been shown to play a key role for multiple individual and organizational outcomes \cite{Granovetter1973-id, Burt2016-as, Reagans2003-gv, uzzi2005collaboration, argote2000knowledge}. Specifically one's neighborhood may consist of several disparate clusters, reflecting their co-workers from different organizational contexts. Prior studies indicate that having diversified connections to bridge different (``well-seperated'') organizational groups allows individuals to access novel and non-redundant information \cite{Ugander2012-yq, De_Choudhury2010-zf, Granovetter1973-id, Burt2016-as, Reagans2003-gv}. 

The contributions of this paper are summarized as follows.
\begin{itemize}
    \item We present the first large-scale, empirical study of the communication network dynamics of new employees in a multinational technology organization in the \textit{remote} work environment. 
    \item Results from our analysis 1) reveal the general time trends of new employee network dynamics across different communication media during remote onboarding, 2) identify the possibly disadvantaged communities in the onboarding network development process, and 3) demonstrate the potential of web-based applications to address employee socialization challenges.
    \item Our work points to several important insights, which may help guide organizational leaders and web application developers on how to quantitatively evaluate and improve the employee socialization experiences at digital workplaces.
\end{itemize}

\section{Hypotheses}

Prior studies suggest that new employee socialization is an assimilation process where network development is arguably a key component \cite{Chen2005-kp, ellis2014new,pike2014new}. This process may unfold over a newcomer's first several months (or years) with an intense ramp-up period at the beginning \cite{ellis2014new}. On the other hand, recent studies suggest that asynchronous communications have largely increased with the broad shift to remote work \cite{Yang2021-rr}. Therefore we speculate the network development patterns of new employees can be different across communication media. Together these inform us the following two hypotheses.

\begin{hypothesis}[\textbf{H1: Time Trends Across Communication Media}]
There exists an onboarding period where communication networks of new employees are continually developing. These network developing trends may vary across different communication media (i.e., remote meetings, emails, and instant messages).
\end{hypothesis}

\begin{hypothesis}[\textbf{H2: New Hires vs. Tenured Employees}]
New employees are less ``connected'' than tenured employees during their onboarding period in the organization (i.e., smaller network size, lower communication intensity, lower structural diversity). However, these differences shall reduce as the tenure of a new employee increases. 
\end{hypothesis}

We speculate there may exist heterogeneity among new employees because of the intrinsic differences of their job duties and social status, thus proposing the following hypothesis.

\begin{hypothesis}[\textbf{H3: Heterogeneity Among New Hires}]
    There exists heterogeneity among new hires (across different managerial positions and different job functions) in how they develop their workplace networks during the onboarding period.
\end{hypothesis}

In addition to the existing investment of human capital on the new employee onboarding process \cite{labuschagne2015onboarding, fagerholm2014role, steinmacher2015social},
with the great workplace digital transformations, we are interested in if new web-based experiences can be enabled to help new employees develop their networks at digital workplaces. 
Previous studies demonstrate that knowledge acquisition and information seeking are key components in the onboarding process and positively relate to new employee socialization outcomes  \cite{ellis2014new,bauer2007newcomer}. 
Therefore we start our exploration by investigating if embedding ``people recommendations'' into organizational knowledge management systems can facilitate new employees in connecting with colleagues they otherwise wouldn't discover (i.e., bridging connections). In this way, knowledge itself serves as an information broker and triggers natural interactions between employees through knowledge sharing, thus improving the network structural diversity.

\begin{hypothesis}[\textbf{H4: Web-Based People Recommendations}]
    New employees who have engaged with the ``knowledge-based people recommendations'' are likely to have more bridging connections.
\end{hypothesis}

We leverage the user engagement data on an enterprise knowledge management product---Microsoft Viva Topics\footnote{https://www.microsoft.com/en-us/microsoft-viva/topics}---to verify the above hypothesis. 
Microsoft Viva Topics is a commercial product which leverages machine learning techniques to compile and surface knowledge within an organization.
It not only surfaces the textual description, but also suggests people in the organizations who specialize in this given knowledge entity. By analyzing how new employees utilize the ``people recommendation'' feature (``Pinned People'' and ``Suggested People'') in this application, we are able to compare the potential differences of their network trends.

\section{Data and Methods}

\subsection{Data Description}\label{sec:data}

We collected a large-scale dataset from Microsoft's full-time employees describing their communication activities from December 2021 to September 2022. The collected data contains anonymized events on major workplace communication platforms: Microsoft Teams for \textit{remote meeting} and \textit{instant messaging}, and Microsoft Outlook for \textit{email services}. Meanwhile, we collected a user profile dataset containing the organizational group, job function, managerial status, location as well as the local team size of each individual user.
Combining these two datasets, we identified two groups of users and conducted statistical analysis on their communication networks.

\begin{itemize}
    \item \textbf{New Hires.} We use the first date of each user's outbound communication activities (i.e., meetings joined, emails sent, instant messages sent) as the estimated employee start date. To mitigate potential biases  introduced by the planned absences (e.g., planned vacations) for these start date estimations, December 2021 is used as a burn-in period to allow users populating in the dataset. After that we identified 11,083 users whose communication activities first started between 2022/01/01 and 2022/03/27 (12 weeks) as the new hire group in our analysis. For each individual user, we then construct weekly snapshots of their communication networks for 24 consecutive weeks from their start week.\footnote{This onboarding window is selected because a) the telemetry data collection has to comply with our data retention policy and thus being restricted to a limited timeframe; and b) this 24-week window is where most onboarding activities take place.}
    \item \textbf{Matched Tenured Employees.} In order to assess the network differences between new hires and tenured employees, we leveraged the above user profile information and identified a matched sample of tenured users in our dataset. For each identified new hire, we matched a tenured user who started before 2022/01/01 on the same profile features including organizational group (63 distinct values), job function (engineer or non-engineer), managerial status (individual contributor or manager), location (headquarter or non-headquarter), local team size (50 distinct values). Only 89 out of 11083 newcomers cannot be matched and are removed in the subsequent analysis. We then extracted their communication networks within exactly the same 24-week observation time window as the one for the paired new hire.
\end{itemize}

\subsection{Constructing Ego Networks}

We're interested in the ego network of each individual in the above identified user groups and how it changes over time. The notation and definitions are formally introduced as the follows.

Given an observation week $t$, a communication medium (meeting, email, or instant messaging), for each individual $k$, we define the undirected \textit{ego network} as $\mathcal{G}^{(t)}_k=\{\mathcal{V}^{(t)}_k, \mathcal{E}^{(t)}_k, \mathcal{W}^{(t)}_k\}$. Here $\mathcal{V}^{(t)}_k=\{k\}\bigcup\mathcal{N}^{(t)}_k$ represents node $k$ as well as its immediate neighbors\footnote{Note $\mathcal{N}^{(t)}_k$ includes all Microsoft full-time employees who have connected with individual $i$ in the given week $t$. The neighboring employees here are not restricted in the user groups identified in \cref{sec:data}.}  ($\mathcal{N}^{(t)}_k$).
$\mathcal{E}^{(t)}_k$ is an undirected edge set which denotes the observed reciprocal connections among nodes in $\mathcal{V}^{(t)}_k$ within the given time window $t$. $\mathcal{W}^{(t)}_k$ represents the corresponding edge weights. 

We first focus on each individual's \textit{one-on-one} networks where $\mathcal{E}^{(t)}_k$ is restricted on the communication interactions with only two participants. 
Following previous studies \cite{kossinets2009origins,De_Choudhury2010-zf}, we define the edge weight $w^{(t)}_{ij}=\sqrt{w^{(t)}_{i\rightarrow j}\times w^{(t)}_{j\rightarrow i}}$ between two individuals $i$ and $j$ as the geometric mean of the number of directed interactions during the observation time period $t$. For example, for each employee's email network, $w^{(t)}_{i\rightarrow j}$ denotes the number of emails that $i$ sent to $j$ in week $t$. To ensure the reciprocity of the observed connections, we define the ego network $\mathcal{G}^{(t)}_k$ comprising the edges with only non-zero weights, i.e., $w^{(t)}_{ij}>0$. We consider an alternative group interaction network for robustness test, which accounts for employees' group communications with less than 10 participants.\footnote{Edge weight ($w^{(t)}_{i\rightarrow j}$) in the group interaction network is relaxed as the number of directed interactions, normalized by the number of recipients within these interactions.}

Following prior studies~\cite{Ugander2012-yq, De_Choudhury2010-zf, easley2010networks}, we consider network metrics to characterize each individual's ego network from three different perspectives: network size, intensity, and structural diversity.

\begin{itemize}
\item \textbf{Number of distinct connections.} This metric is defined as the total number of employees connected to the focal individual $k$ within the observation time window $t$, which directly measures the \textit{size} of one's neighborhood, i.e., $|\mathcal{N}^{(t)}_k|$.

\item \textbf{Sum of edge weights.} We sum over the weights of all edges connected to the ego node $k$ to assess its overall communication \textit{intensity}, i.e., $W_k^{(t)}=\sum_{j \in \mathcal{N}^{(t)}_k} w^{(t)}_{ij}$. Note the units of edge weights vary in communication media, where the weights for meeting, email, and IM networks are calculated based on the number of joint meetings, emails exchanges, and joint chat sessions respectively.

\item \textbf{Number of ego components.} We leverage the number of ego components ($C_k^{(t)}$) to measure how much each individual $k$ is bridging surrounding communities. This measure is defined as the number of connected components that remain in the ego network $\mathcal{G}^{(t)}_k$ when the focal node $k$ as well as its incident edges are removed. This metric is frequently adopted in prior network science studies to describe the \textit{structural diversity} of one's ego network \cite{Ugander2012-yq, De_Choudhury2010-zf}.

\end{itemize}

\subsection{Organizational Attributes}\label{sec:attributes}
In addition to the above network metrics, we also consider the following organizational attributes as control variables or predictors in our analysis.

\begin{itemize}
\item \textbf{Organizational Group.} This attribute represents the highest-level organizational group within Microsoft that each individual employee belongs to. In our dataset, it is defined as the parent nodes one step below the CEO (the root node) in the formal organizational chart.
\item \textbf{Job Function.} We consider a binary variable $x^{\mathit{(eng)}}$ to represent if an individual is at the engineering function role, i.e., engineer ($x^{\mathit{(eng)}}=1$) versus non-engineer ($x^{\mathit{(eng)}}=0$).
\item \textbf{Managerial Status.} This attribute is defined as an indicator $x^{\mathit{(mngr)}}$ to reflect if the employee is in the management position, i.e., manager ($x^{\mathit{(mngr)}}=1$) versus individual contributor ($x^{\mathit{(mngr)}}=0$).
\item \textbf{Location.} We introduce a binary variable to represent if each employee's office (whether it is a home office or not) is located in the same metropolitan area as Microsoft's headquarter office.
\item \textbf{Local Team Size.} We calculate the number of employees who directly report to the same direct manager as the target individual and use this variable to assess the number of peer co-workers in one's direct team.
\end{itemize}

\begin{table*}[t]
    \centering
    \small
    \begin{tabular}{lllll}
    \toprule
        \textbf{Hypothesis} & \textbf{Model} & \textbf{Scope} & \textbf{Variable(s) of Interests} & \textbf{Control Variables} \\
        \midrule
        \textbf{H1} & \textbf{M1} & New hires only & Onboarding Week & \makecell[l]{Organizational Group, Job Function, Managerial Status,\\Location, Local Team Size} \\
        \midrule
        \textbf{H2} & \textbf{M2} & \makecell[l]{New hires +\\tenured employees}& Tenure Group & \makecell[l]{Organizational Group, Job Function, Managerial Status,\\Location, Local Team Size} \\
        \midrule
        \textbf{H3} & \textbf{M3.a} & New hires only &\makecell[l]{Managerial Status, Onboarding Week\\Managerial Status $\times$ Onboarding Week} &  \makecell[l]{Organizational Group, Job Function,\\Location, Local Team Size} \\
        \midrule
        \textbf{H3} & \textbf{M3.b} & New hires only &\makecell[l]{Job Function, Onboarding Week\\Job Function $\times$ Onboarding Week} &  \makecell[l]{Organizational Group, Managerial Status,\\Location, Local Team Size} \\
        \midrule
        \textbf{H4} & \textbf{M4} & New hires only &\makecell[l]{People Recommendation, Onboarding Week\\People Recommendation $\times$ Onboarding Week}  & \makecell[l]{Organizational Group, Job Function, Managerial Status,\\Location, Local Team Size} \\
         \bottomrule
    \end{tabular}
     \Description[Summary of the model designs, where each model is conducted on meeting, email and IM networks separately.]{Summary of the model designs, where each model is conducted on meeting, email, and IM networks separately.}
    \caption{Summary of the model designs, where each model is conducted on meeting, email, and IM networks separately.}
    \label{tab:models}
\end{table*}

\subsection{Models}\label{sec:models}

As described in \cref{sec:data}, we calculate the network statistics of each identified new hire and the matched tenured employee's weekly ego networks across different communication platforms over 24 consecutive weeks. To examine our proposed hypotheses, we follow the mixed effect framework and perform a series of longitudinal ordinary least squares (OLS) regressions. The regression framework can be described as below
\begin{equation}
    y_{k}^{(t)} \sim \mathit{variables\ of\ interests} + \mathit{control\ variables} + \eta_t + \epsilon_k . \label{eq:MixedLM}
\end{equation}
For each individual $k$ in the observation week $t$, the dependent variable ($y_{k}^{(t)}$) encodes one of the following three network metrics---the number of distinct connections ($|\mathcal{N}^{(t)}_k|$), the sum of edge weights ($W^{(t)}_k$), and the number of ego components ($C^{(t)}_k$). In order to control for the potential time unit-specific bias (e.g., the lack of activeness during the holiday week), a time-dummy variable $\eta_t$ is included to capture these time fixed effects. To model the potential variations across different individuals, $\epsilon_k$ is introduced to capture the random effects. 
For each individual new hire $k$, we also introduce \textit{Onboarding Week} as a numeric time-varying variable. 
For each new hire $k$ within the given observation week $t$, the corresponding \textit{Onboarding Week} can be defined as $t - \tilde{t}_{k}$, where $\tilde{t}_{k}$ denotes the start week of the individual $k$.
We discuss how we operationalize this framework to examine each hypothesis as the following. A summary of the detailed model designs are included in \cref{tab:models}.

To test whether new hire's network connectivity metrics increase over the monitored 24-week onboarding period and how these metrics differ across various communication platforms (\textbf{H1}), we perform the regression analysis on new hires' meeting, email as well as instant messaging (IM) ego networks. We examine the \textit{Onboarding Week} as the variable of interests in \cref{eq:MixedLM}, and the organizational attributes (as defined in \cref{sec:attributes}) are utilized as control variables (\textbf{M1}). The estimated coefficient of \textit{Onboarding Week} allows us to assess the direction as well as the speed of network connectivity changing over the new hire onboarding period.

To test the potential differences between new hires and tenured employees' networks (\textbf{H2}), we construct a combined dataset with both new hires as well as the matched tenured employees. We introduce a binary variable \textit{Tenure Group} $x_k^{\mathit{(tenure)}}$, where $x_k^{\mathit{(tenure)}}=1$ indicates the individual $k$ belongs to the new hire user group and $x_k=0$ otherwise. We use all organizational attributes as control variables and test the effect of \textit{Tenure Group} (\textbf{M2}). The estimated effect then corresponds to the average difference between the new hire group and the tenured employee group within the observation time window.

To examine the potential differences of network connectivity metrics as well as their growing speeds between new managers and new individual contributors (\textbf{H3}), we consider the \textit{Managerial Status}, \textit{Onboarding Week} as well as their \textit{interaction} as the variables of interests (\textbf{M3.a}). All other organizational variables are used as the control variables. In this model, the estimated effect of the \textit{Managerial Status} reflects the average difference between new managers and new individual contributors while the interaction between \textit{Managerial Status} and \textit{Onboarding Week} indicates the difference of growth rates of their networks over time. By switching the \textit{Managerial Status} variable and the \textit{Job Function} variable, we perform a similar regression analysis to examine the potential differences between engineers and non-engineers during their 24-week onboarding time (\textbf{M3.b}).

To test whether the adoption of the knowledge-based ``people recommendation'' application associates with a new hire's networking behavior (\textbf{H4}), we introduce an additional binary variable \textit{People Recommendation} $x_k^{\mathit{(rec.)}}$ to categorize new hires based on their engagement behavior on these recommendations, where $x_k^{\mathit{(rec.)}}=1$ indicates the individual $k$ has clicked or viewed the ``Pinned People'' and ``Suggested People'' feature in Viva Topics within the tracked 24-week onboarding period and $x_k^{\mathit{(rec.)}}=0$ otherwise. Similar to the previous setup, we test \textit{People Recommendation}, \textit{Onboarding Week} as well as their first-order \textit{interaction} and include all organizational attributes as control variables (\textbf{M4}). By testing \textit{People Recommendation}, we aim to identify its potential correlation with one's network connectivity metrics. By testing the interaction term (\textit{People Recommendation} $\times$ \textit{Onboarding Week}), we explore its potentials on one's network connectivity growth rate during the onboarding period.

\section{Results and Discussion}

\begin{figure*}[t]
    \centering
    \includegraphics[width=\textwidth]{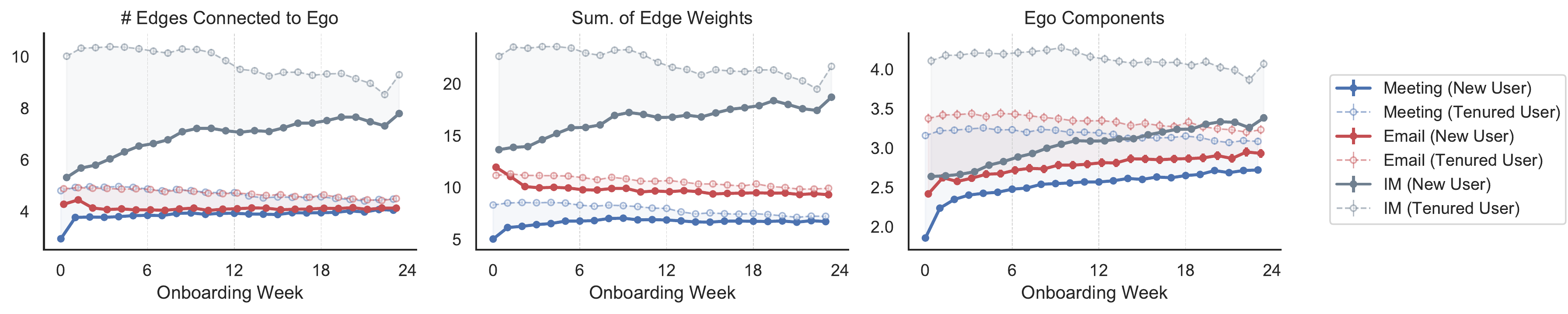}
    \Description[Time trend comparisons on different communication media---meeting, email and instant messaging (IM)]{Time trend comparisons on different communication media---meeting, email, and instant messaging (IM)} 
    \caption{Time trend comparisons on different communication media---meeting, email, and instant messaging (IM)---which are highlighted using different colors. Data points represent the means of network statistics across users in the given observation week and error bars indicate the 95\% confidence intervals (covered by the markers in some places). The differences between new hires and tenured employees on each platform are shaded using its corresponding color. }
    \label{fig:multilayer_trends_1on1}
\end{figure*}

\subsection{Communication Trends}\label{sec:trends}
We start by presenting model-free time trends for different network metrics. These trends describe how the communication networks of new hires change across different communication media over the 24-week onboarding period. By comparing these new hires' trends to the matched tenured employees within the same observation window, we highlight how new hires assimilate their communication behavior to tenured employees over time. Then we conduct a series of statistical regression analyses (as described in \cref{sec:models}) to formally examine our hypotheses.

\begin{figure}
\centering
    \includegraphics[width=\linewidth]{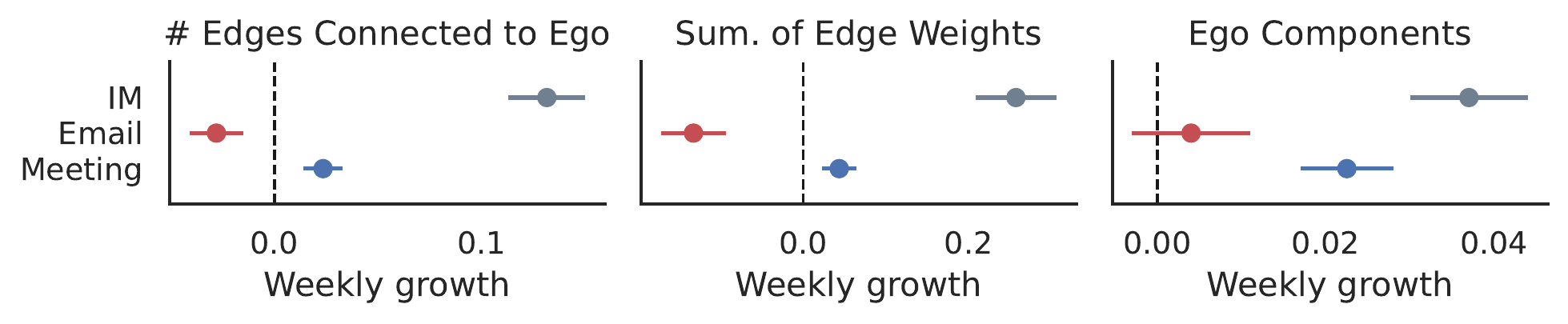}
    \caption{Weekly growth rates (with 95\% CIs) of network statistics (M1).}\label{fig:CI_NA_OnboardingWeek_overall_1on1}
    \Description[Weekly growth rates (with 95\% CIs) of network statistics (M1).]{Weekly growth rates (with 95\% CIs) of network statistics (M1).}
\end{figure}

\paragraph{Communication Media} 
\cref{fig:multilayer_trends_1on1} shows the average weekly time series for network connectivity statistics on different communication platforms across different tenure groups (new hires versus matched tenured employees). Focusing on the trends of new hires' network statistics, we observe different onboarding patterns from this figure with respect to the use of communication media (\textbf{H1}). We find new hires' one-on-one \textit{meeting} and \textit{instant messaging} (IM) networks are overall expanding over the 24-week onboarding period with respect to their network size, intensity, and structural diversity, where the size and the intensity of meeting networks rapidly converge in the first few weeks while the IM networks are continually growing on all three network metrics. On the other hand, we notice new hires generally started with larger and more intense \textit{email} networks and these metrics quickly dropped in their first few onboarding weeks.\footnote{A possible explanation for this observation is in the first few weeks, newcomers often engage with ad-hoc activities requiring formal written communications (e.g., introduction emails), thus leading to the bump in their email network trends.} To validate these observations, we test the \textit{Onboarding Week} variable in our statistical regression model (\textbf{M1}) over the 24-week onboarding window and present the corresponding coefficients with the 95\% confidence intervals (CIs) in \cref{fig:CI_NA_OnboardingWeek_overall_1on1}. We find a statistically significant weekly growth of new hires' IM networks on all three network metrics over the 24-week observation time ($+0.131$ per week on network size, $p<0.001$; $+0.257$ per week on network intensity, $p<0.001$; $+0.037$ per week on structural diversity, $p<0.001$). These effects on IM networks appear to be significantly larger than those on other communication media, which confirms the fast-growing patterns we observed about new hires' networking behavior via instant messaging.

In addition to our observations about communication media choice among new hires, we also notice a strong preference from tenured employees on one-on-one communications over instant messages (i.e., larger communication network size, intensity, and diversity). Although employees' choices on communication media can be affected by various factors, including both firm-wide guidelines and personal inclinations, this preference for instant messages is in fact consistent with the findings from recent studies where asynchronous and informal communications have increased among information workers with the shift to remote work \cite{Yang2021-rr}. 
The growing trend of communicating through instant messages may also reflect new hires' assimilation process, where they gradually adopt the communication media choice from tenured employees in the remote and hybrid work environment. 
\raggedbottom

\begin{figure}
    \centering
    \includegraphics[width=\linewidth]{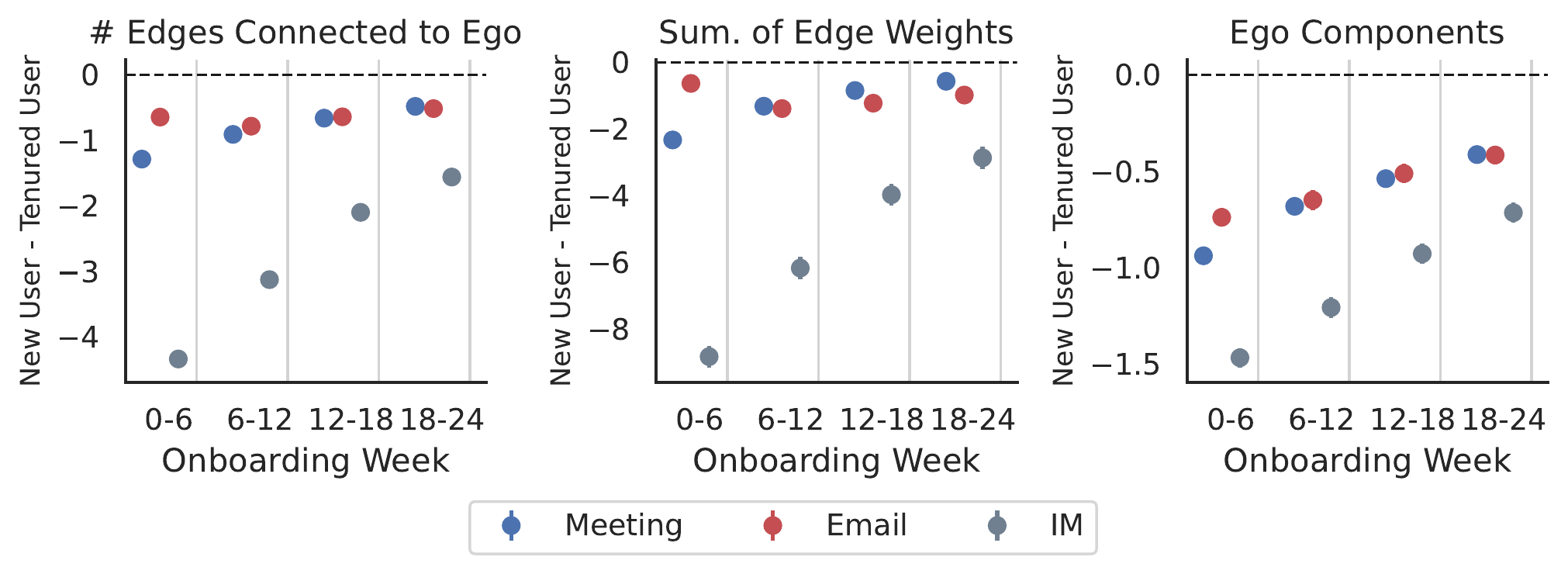}
    \Description[Differences of network statistics (with 95\% CIs) between new hires and tenured employees within each 6-week testing window (\textbf{M2}).]{Differences of network statistics (with 95\% CIs) between new hires and tenured employees within each 6-week testing window (\textbf{M2}).} 
    \caption{Differences of network statistics (with 95\% CIs) between new hires and tenured employees within each 6-week testing window (\textbf{M2}).}
    \label{fig:isNewhire_trends_1on1}
\end{figure}

\paragraph{New Hires vs. Tenured Employees}
\cref{fig:multilayer_trends_1on1} also reveals clear differences between the communication networks of new hires and tenured employees (\textbf{H2}). Compared to new hires, tenured employees generally connect with more colleagues every week (larger network size), communicate more intensively with other colleagues (larger edge weights), and maintain more diversified connections within the company (higher structural diversity) across all communication media. Despite their time trend differences on different communication media in the first few onboarding weeks, we observe new hires' network statistics eventually trend towards the states of tenured employees at the later stage of the tracked 24-week onboarding period. We investigate this pattern by conducting finer-grained statistical regressions to test the \textit{Tenure Group} variable (\textbf{M2}) on every 6 onboarding weeks and report the results in \cref{fig:isNewhire_trends_1on1}. We notice the network statistics of new hires and tenured employees are significantly different ($p<0.001$), while these differences are diminishing as the tenure of a new employee increases.

Through the above analysis, the socialization aspect of the organizational assimilation process can be reflected on these time trends of new hires' communication networks, where their network size, intensity, and diversity are approaching the states of tenured employees in the organization. Regardless we still find statistically significant gaps on these metrics between different tenure groups at the latest state of the tracked onboarding period, implying it may take longer than 24 weeks (6 months) to have new employees fully ``onboarded'', i.e., statistically indistinguishable from tenured employees on network size, intensity and diversity.

\subsection{Heterogeneity Among New Employees} \label{sec:heterogeneity}
While we observe a general trend that new hires are assimilating their networks to those of tenured employees,
we hypothesize there may exist heterogeneity across new employees due to the fundamental differences of their job duties, social status, and day-to-day workflows~\cite{van1976determinants,ruchinskas1983communicating,diesner2005exploration,Romero2016-ug} (\textbf{H3}). We conduct descriptive analysis and statistical regressions to investigate their heterogeneity on the following two dimensions: managerial position (manager vs. individual contributor) and job function (engineer vs. non-engineer).

\begin{figure}
    \centering
\begin{subfigure}[b]{\linewidth}
    \centering
    \includegraphics[width=0.9\linewidth]{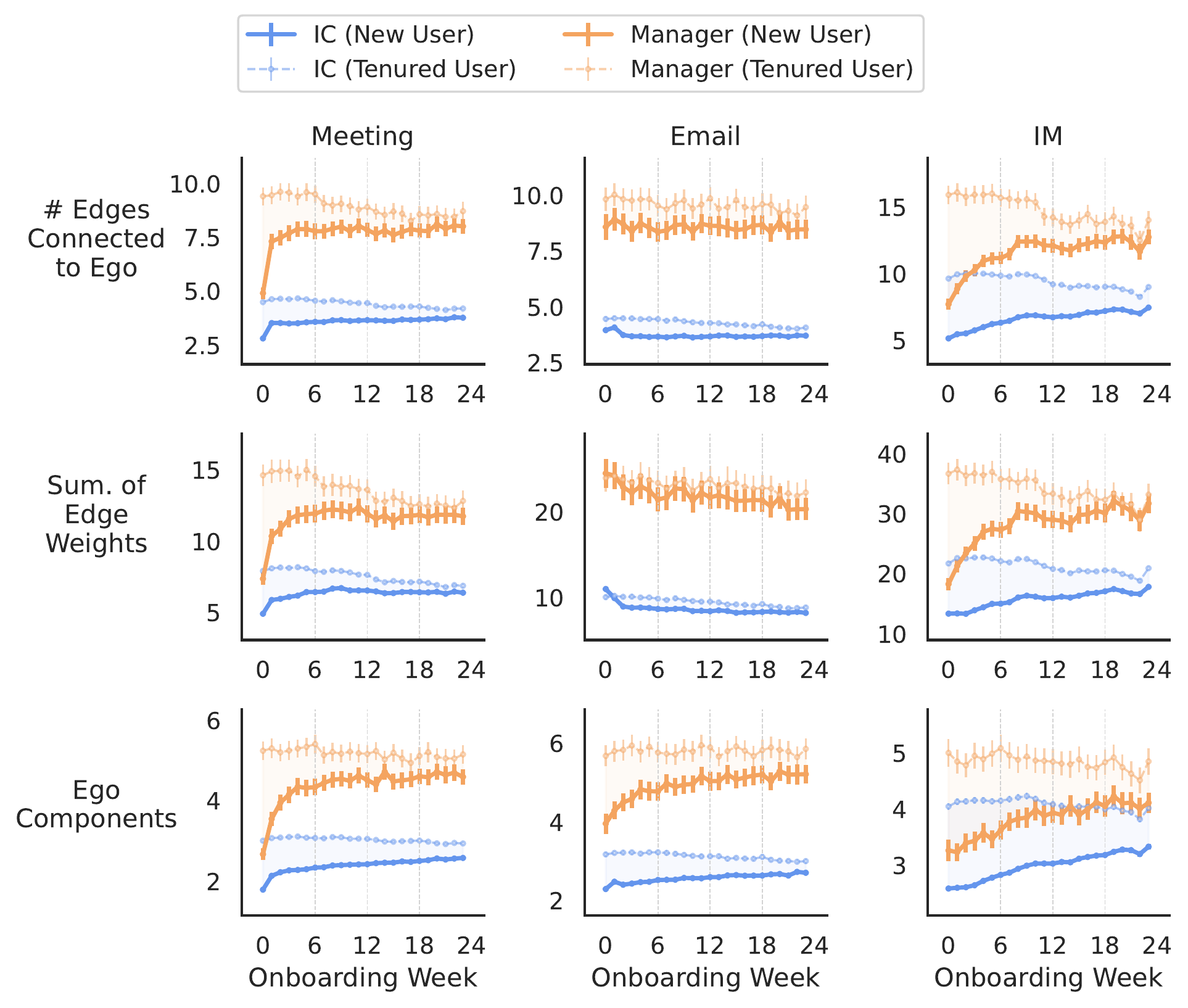}
    \caption{Heterogeneity w.r.t. Managerial Status (Manager vs. Individual Contributor).}
    \label{fig:isManager_trends_1on1}
\end{subfigure}

\begin{subfigure}[b]{\linewidth}
    \centering
    \includegraphics[width=0.9\linewidth]{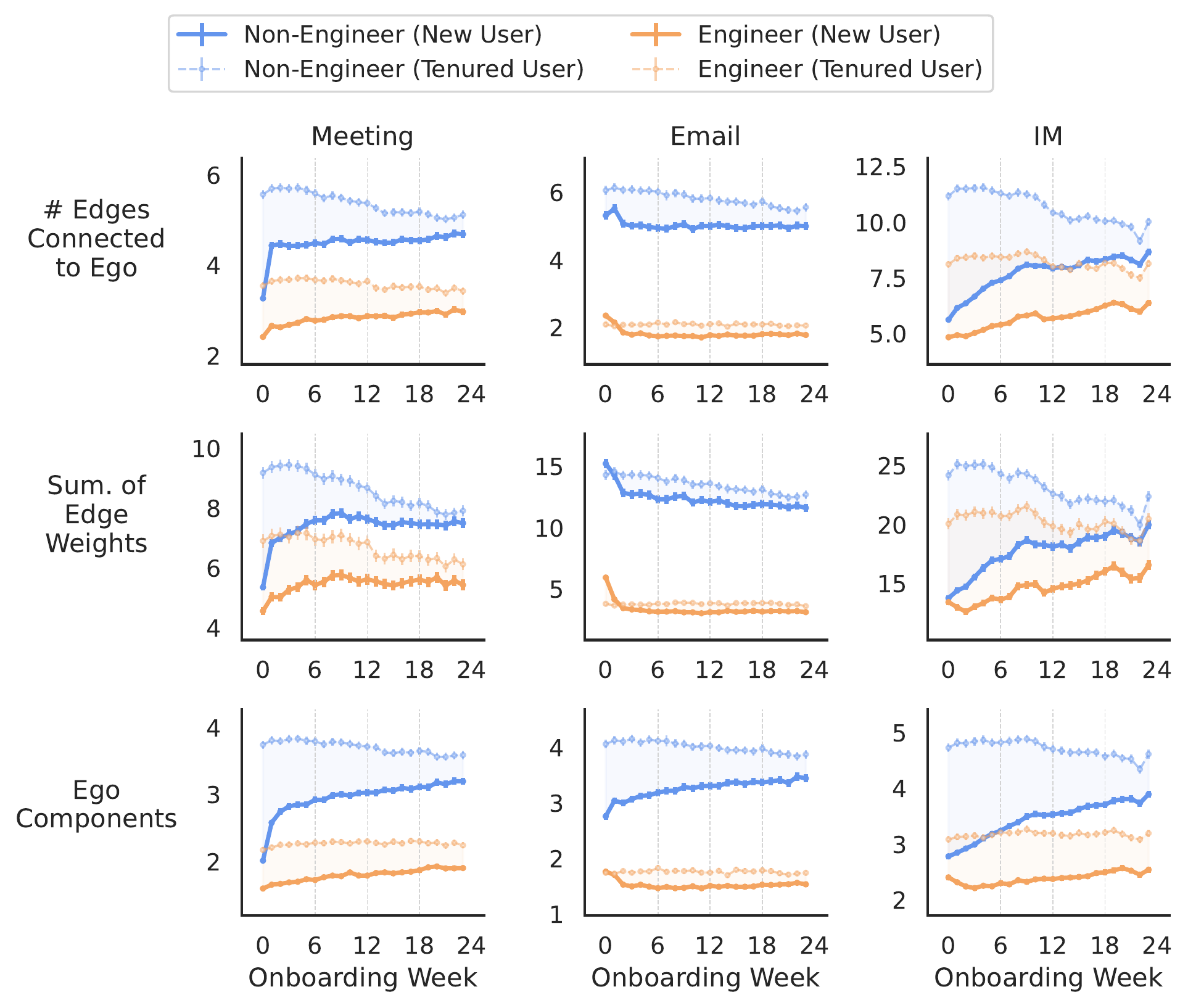}
    \caption{Heterogeneity w.r.t. Job function (Engineer vs. Non-Engineer).}
    \label{fig:isSWE_trends_1on1}
\end{subfigure}

\caption{Time trend comparisons across different new hire groups on their meeting, email, and instant messaging networks (average network statistics with 95\% CIs).}

\Description[Time trend comparisons across different new hire groups on their meeting, email, and instant messaging networks (average network statistics with 95\% CIs).>]{Time trend comparisons across different new hire groups on their meeting, email, and instant messaging networks (average network statistics with 95\% CIs).1) Heterogeneity w.r.t. Managerial Status (Manager vs. Individual Contributor).2)Heterogeneity w.r.t. Job function (Engineer vs. Non-Engineer).} 
    \label{fig:worker_heterogeneity_trends_1on1}
\end{figure}

\begin{figure}
        \centering
\begin{subfigure}[b]{\linewidth}
    \centering
    \includegraphics[width=\linewidth]{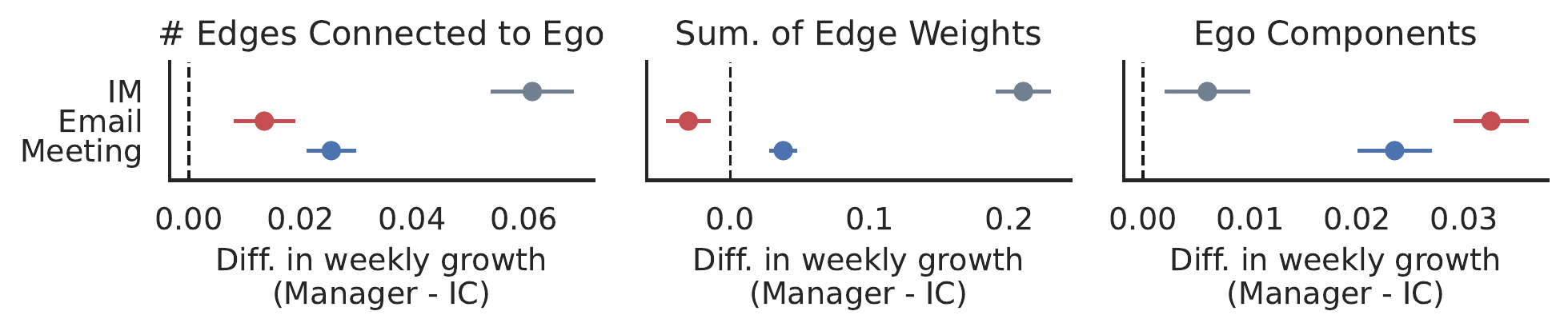}
    \caption{Manager vs. Individual Contributor.}\label{fig:CI_isManager_OnboardingWeek_overall_1on1}
\end{subfigure}

\begin{subfigure}[b]{\linewidth}
    \centering
    \includegraphics[width=\linewidth]{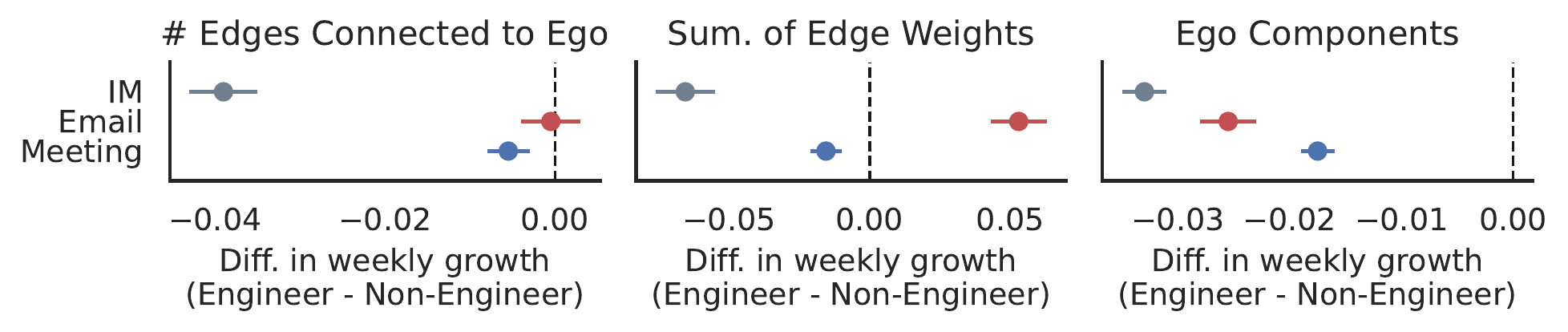}
    \caption{Engineer vs. Non-Engineer.}\label{fig:CI_isSWE_OnboardingWeek_overall_1on1}
\end{subfigure}

\caption{Heterogeneity of network growth rates (with 95\% CIs) across different new hire groups.}
 \Description[Heterogeneity of network growth rates (with 95\% CIs) across different new hire groups.]{Heterogeneity of network growth rates (with 95\% CIs) across different new hire groups.Manager vs. Individual Contributor. Engineer vs. Non-Engineer.}
\end{figure}

\paragraph{Managers vs. Individual Contributors} Among the 11,083 new hires we identified, we observe 6.37\% of them are managers and 93.63\% of them are individual contributors. 
\cref{fig:isManager_trends_1on1} presents the weekly trends of network statistics for employees with different managerial statuses, where we observe a distinct pattern that compared to individual contributors (ICs), managers connect with more employees in the organization and their communication networks tend to be more intense and more structurally diverse. Note these differences are consistent across both new hires and tenured employees, reflecting the intrinsic nature of their role types: manager roles are usually obliged to conduct people coordination while IC roles are often expected to focus on ``hands-on'' work practices. We further investigate the differences among new hires and notice that managers' communication networks generally grow and stabilize faster than individual contributors. To quantify the differences of their network growth rates, we test the first order interaction between \textit{Managerial Status} and \textit{Onboarding Week} (\textbf{M3.a}) and report these effects in \cref{fig:CI_isManager_OnboardingWeek_overall_1on1}. We find the growth rates on various network metrics are significantly different across managers and ICs. In particular, we find new managers' networks grow significantly and consistently faster than ICs on their structural diversity metrics across all communication media ($+0.023$ per week on remote meeting, $p<0.001$; $+0.032$ per week on email, $p<0.001$; $+0.006$ per week on instant messaging, $p=0.002$).

A possible explanation here is in comparison with individual contributors, extensive communications and coordinations are required in managers' daily work (e.g., host routine one-on-one meetings with their direct reports, maintain frequent check-ins with representatives from other groups), which creates both demands and opportunities for new hires who are in manager roles to rapidly develop their networks. On the contrary, individual contributors may be put on a disadvantaged position at their onboarding process regarding their social network building. Such a disadvantage can be worsen with the current shift to remote work due to the absence of spontaneous in-person connections at workplace (e.g., impromptu lunches, hallway conversations) \cite{Yang2021-rr,Bojinov2021-xy,Cooper2002-nc,nardi2002place,Golden2006-zp}.

\paragraph{Engineers vs. Non-Engineers.} We observe that 35.67\% of our identified new hires are engineers while the other 64.33\% are in non-engineering roles (e.g., program managers, sales representatives). 
\cref{fig:isSWE_trends_1on1} presents the weekly trends of network statistics for employees with different job functions. Similarly we observe a clear pattern that engineers connect with fewer employees, with less total communication intensity, and their networks tend to be less structurally diverse compared to non-engineers. We also notice that new engineers' communication networks in general expand slower than non-engineering new hires, especially in terms of their network structural diversity. By testing the interaction between \textit{Job Function} and \textit{Onboarding Week} through regression analysis (\textbf{M3.b}), we are able to validate these observations (as shown in \cref{fig:CI_isSWE_OnboardingWeek_overall_1on1}). Notably we find engineering new hires' network growth rates on structural diversity are significantly lower than non-engineers across all communication platforms ($-0.017$ per week on remote meeting, $p<0.001$; $-0.026$ per week on email, $p<0.001$; $-0.033$ per week on instant messaging, $p<0.001$).

Relative to non-engineers (e.g., program managers, sales representatives), we conjecture new employees in the engineering role may be less in need of large and diverse communication networks in order to perform their daily jobs tasks, therefore lacking opportunities to organically grow their networks (especially on weak and bridging ties, which is reflected on the number of ego components).

\subsection{Opportunities for Web-Based Applications}

The above results reveal that it may take a long time ($>24$ weeks) to fully onboard new employees from the socialization perspective, and certain groups of new hires could be at a disadvantaged position in this process because of the limited opportunities to expand networks when performing day-to-day requirements of the job. Nevertheless, workplace digital transformations during COVID-19 pandemic also introduced opportunities to address this connectivity crisis by enabling new experiences on web-based digital applications. In addition to provide disruptive interventions such as intentionally hosting virtual ``watercooler'' sessions, we also investigate if web-based nudges can be seamlessly embedded into an employee's day-to-day workflows, thus triggering natural interactions with colleagues whom they otherwise wouldn't connect with and improving their network structural diversity.

\paragraph{Knowledge-Based People Recommendation}
Given knowledge acquisition is a critical component of the onboarding process for information workers, we start by investigating if providing ``people recommendations'' in organizational knowledge management systems can help new employees connect with colleagues otherwise they wouldn't know. Specifically, we test if there are differences of network dynamics between new employees who have or not engaged with (viewed or clicked) these ``people recommendations'' (i.e., the ``Pinned People'' and ``Suggested People'' feature in Microsoft Viva Topics) during their onboarding period. These statistical tests (\textbf{M4}) are performed on every 6 onboarding weeks and the results are presented in \cref{tab:topics_diff_ego_components_overall}, where we observe statistically significant differences regarding network diversity and the weekly growth rate of this metric in the first 12 onboarding weeks, indicating the potential positive effects of presenting ``people recommendations'' on new employees' network diversity. However we find such effects become less significant in the late onboarding stage. A possible explanation is new hires in our dataset are normally leverage this organizational knowledge base (Microsoft Viva Topics) as an onboarding tool to explore the organizational knowledge taxonomy in their early days. As the onboarding process progresses, their workflows start shifting from exploring organizational knowledge to executing job tasks. This observation also implies future research directions on fully investigating potential canvases and approaches to present ``people recommendations'' based on the employee's working context.

\begin{table}[]
    \centering
    \small
\begin{subtable}[]{\linewidth}
\centering
\begin{tabular}{clll}
\toprule
\makecell[c]{Onboarding Week} &  Email &  IM  &  Meeting  \\
\midrule
0-6   &            0.194*** &  0.141*** &              0.028 \\
6-12  &            0.200* &  0.033 &             -0.024 \\
12-18 &            0.120 &  0.140 &              0.051 \\
18-24 &            0.242 &   0.097 &              0.088 \\
\bottomrule
\end{tabular}
   \caption{Differences of network structural diversity (*$p<0.05$, **$p<0.01$, **$p<0.001$).}
    \label{tab:topics_diff_ego_components}
\end{subtable}\\

\begin{subtable}[]{\linewidth}
    \centering
    \small
\begin{tabular}{clll}
\toprule
\makecell[c]{Onboarding Week} &  Email  &  IM &  Meeting \\
\midrule
0-6   &                 0.026** &              0.018** &                   0.029*** \\
6-12  &                 0.013 &              0.026*** &                   0.019** \\
12-18 &                 0.015 &              0.011 &                   0.009 \\
18-24 &                 0.005  &              0.011 &                   0.005 \\
\bottomrule
\end{tabular}
    \caption{Differences of the weekly growth of network structural diversity (*$p<0.05$, **$p<0.01$, **$p<0.001$).}
    \label{tab:topics_diff_ego_components_growth}
\end{subtable}
\caption{The relationship between new hires' engagements with the ``Pinned People'' and ``Suggested People'' feature in Microsoft Viva Topics and their network structural diversity (\textbf{M4}), i.e., the number of ego components.}\label{tab:topics_diff_ego_components_overall}
\end{table}

\subsection{Additional Analysis on Group Interaction}

Our analysis thus far has considered the one-on-one communication patterns of newcomers during the virtual onboarding period. We also conducted a separate analysis for group communications (with less than 10 participants in each group interaction).  Our findings remain consistent on the significant gaps of all network statistics between new hires and tenured employees after the six-month onboarding phase, the heterogeneity across different managerial positions and job functions, and the positive association between the usage of web-based people recommendations and the network structural diversity of new hires. 
More details are included in the appendix.

\section{Related Work}
\paragraph{Remote work}
Remote work is an important topic to study across academia and industry. Remote work-related studies have covered various areas, such as communication challenges~\cite{Rodeghero2021-rr,he2014qualitative}, well-being~\cite{ butler2021challenges, michel2021flattening}, management~\cite{teevan2020new}, etc. While most work focuses on distributed work or remote work due to geographical distance~\cite{hinds2002distributed, Koehne2012-id, nardi2002place, Leonardi2010-jn, Britto2018-at}, ours  focuses on virtual trending caused by COVID-19. Related to our study, a stream of research emphasizes the impact of remote work on organizational workers’ connectivity in general. Most of them feature the strength of existing or new interpersonal ties~\cite{Yang2021-rr,Yang2022-ei,carmody2022effect,Wu2021-nz}. For example, scholars find that firm-wide remote work triggered by COVID leads to a more siloed network among information works with the decrease weak and bridging ties~\cite{Yang2021-rr} and in general, ties are more difficult to form in remote work settings~\cite{carmody2022effect}.

This could have significant ramifications, as connections between workers in in-person settings have been associated with organizational outcomes such as building successful collaborations~\cite{kabo2014proximity}, relationships, and with knowledge transfer~\cite{allen2015effective}.  Organizational knowledge is inherently social, and is dependent on shared language, narratives, identification, norms, commitment and trust~\cite{ackerman2004sharing}. Additionally, the ability to build connections enables workers to self-organize and share expertise~\cite{ackerman2004sharing}. Research prior to the pandemic indicates that while face-to-face conversation is preferred for relationship building, ideation, and problem solving, communication technologies such as IM and social network sites perform complementary purposes including building and maintaining connections~\cite{turner2010exploring}. In this paper, we focus on virtual onboarding rather than studying full-time tenured organizational workers, and study how use of a range of technologies, including one that highlights other organization members to the user, are related to workers' connectivity.

\paragraph{Virtual onboarding}
While traditional, in-person onboarding  has been frequently studied in the context of organization outcomes~\cite{Chen2005-kp, Cable2013-kj, Fagerholm2014-dv, Steinmacher2019-ue, Snell2006-qm}, virtual onboarding is still an open area. Most studies focus on virtual onboarding challenges related to specific job functions such as engineering~\cite{Rodeghero2021-rr, Dominic2020-qn, Britto2018-at, Fagerholm2014-dv}. For example, a very relevant qualitative study by Rodeghero et al. indicates that remote work harms social connection, but changing the mode of interaction (such as turning on the camera during meetings) could potentially improve socialization~\cite{Rodeghero2021-rr}. Previous work also discusses AI tools (e.g., voice assistant tools)~\cite{upadhyay2018applying, dominic2020onboarding}, or human effort (e.g., hosting virtual watercooler events~\cite{bojinov2021virtual}, or providing checklists or mentorship~\cite{Fagerholm2014-dv, Canfora2012-fx}) that can be used to improve onboarding. Our work further explores the remote onboarding process in the context of connectivity and provides potential web solutions that connect people with the expert knowledge needed to improve employee connectivity. 

\section{Conclusion and Future Work}

In this paper, we investigated the dynamics of new employees' workplace communication networks throughout a 24-week onboarding period. We also explored the potential web-based applications to address the socialization challenge and summarize our findings as follows.
\begin{itemize}
    \item While new hires are less connected with colleagues in the organizations compared to tenured employees in the 24-week onboarding period (smaller network size, lower communication intensity and lower structural diversity), their networks are gradually expanding over time.
    \item Time trends of new hires' network statistics vary across different communication media, managerial positions, and job functions. We observed that new employees, whose day-to-day job tasks are not necessarily reliant on extensive and diverse collaborations (e.g., individual contributors, software engineers), can be put into a disadvantaged position regarding organizational socialization in the onboarding process.
    \item We explored the opportunities for web-based applications to address the new employee socialization challenges. Early evidence revealed that presenting ``people recommendations'' within employees' workflows may benefit their workplace network development during the early onboarding period (first 12 weeks).
\end{itemize}

This study, to the best of our knowledge, presents the first large-scale empirical investigation on communication network dynamics of new employees in the technology industry.
This study is not without limitations. 
First, given the availability and retention policy of the dataset, our study was restricted to a 24-week onboarding window. This constraint can be further relaxed if more data become available since we empirically revealed that it may take longer than 24 weeks to observe new hires' network statistics fully converge to the states of tenured employees. In this way, one can evidently document the convergence time of new employees with different job roles and empirically understand the dynamics of new hires' networks in the long term. These quantitative analyses can also be complemented with qualitative studies to deeply understand new employees' networking behavior.
Second, new employees studied in this work mostly followed the remote onboarding guidelines within the organization, thus limiting our scope within the remote work context. We acknowledge that the current and future hybrid work policies could impact new employees' socialization experiences. Therefore a future research direction is to extend the empirical analysis and investigate the potential onboarding challenges as well as opportunities induced by the emerging trend of hybrid work.
Third, our study is limited to a single firm and, for the most part, information workers. A future direction could be expanding the analysis scope and investigating the potential heterogeneity regarding new employee network dynamics across different firms and industries.

We quantitatively described how new employees develop their workplace network over a 6-month onboarding period in this study. Our results confirmed the struggle and the need of ``staying connected'' from new employees in the remote work environment. Our study revealed there is no single network development path for all job functions. We also want to highlight that overlooking the socialization needs from disadvantaged employee groups may result in negative effects on individuals, organizations, and social fairness. In addition to the investments of human capitals on new employee onboarding experiences, we demonstrated that recent workplace digital transformations have introduced possibilities for web-based technologies to address the new employee socialization challenges. Here, we’d like to call for both human capital investments (e.g., host social events) and technical solutions (e.g., people recommendations, virtual “watercooler” sessions) on addressing the socialization needs of new hires. Network metrics in this study can also be leveraged as a toolkit to identify/self-identify newcomers who may need help. We hope our study can shed some lights on these opportunities and encourage future research efforts from the web community to keep advancing digital workplaces.

%\newpage
\balance
\bibliographystyle{ACM-Reference-Format}
\bibliography{sample-base}

\newpage

\appendix

\section{Appendix}

\begin{table}[h]
    \centering
    \small
    \begin{subtable}[]{\linewidth}
            
\begin{tabular}{llrrr}
\toprule
        & &  \makecell[c]{Num. of \\connections} &  \makecell[c]{Sum. of \\edge weights} &  \makecell[c]{Num. of \\ego components} \\
                \midrule
\multirow{3}{*}{Meeting} & Min &                  1.00 &                 1.00 &           1.00 \\
        & Max &                 38.00 &                79.00 &          26.00 \\
        & Mean &                  3.87 &                 6.67 &           2.53 \\
\midrule
\multirow{3}{*}{Email} & Min &                  1.00 &                 1.00 &           1.00 \\
& Max &                 79.00 &               168.89 &          57.00 \\
        & Mean &                  4.13 &                 9.85 &           2.77 \\
        \midrule
\multirow{3}{*}{IM} & Min &                  1.00 &                 1.00 &           1.00 \\
        & Max &                138.00 &               220.45 &          77.00 \\
        & Mean &                  6.94 &                16.56 &           3.04 \\
\bottomrule
\end{tabular}
\caption{One-on-one networks.}
    \label{tab:stats-1-on-1}
    \end{subtable}\\
    
    \begin{subtable}[]{\linewidth}
\begin{tabular}{llrrr}
\toprule
        & &  \makecell[c]{Num. of \\connections} &  \makecell[c]{Sum. of \\edge weights} &  \makecell[c]{Num. of \\ego components} \\
                \midrule
\multirow{3}{*}{Meeting} & Min &                  1.00 &                 1.00 &           1.00 \\
        & Max &                 100.00 &                118.18 &          21.00 \\
        & Mean &                  12.68 &                 10.99 &           1.76 \\
\midrule
\multirow{3}{*}{Email} & Min &                  1.00 &                 1.00 &           1.00 \\
& Max &                 124.00 &               243.51 &          86.00 \\
        & Mean &                  7.56 &                 14.11 &           2.35 \\
        \midrule
\multirow{3}{*}{IM} & Min &                  1.00 &                 1.00 &           1.00 \\
        & Max &                142.00 &               236.96 &          74.00 \\
        & Mean &                  9.76 &                20.53 &           2.59 \\
\bottomrule
\end{tabular}
\caption{Group interaction networks.}
    \label{tab:stats-group}
    \end{subtable}
    \caption{Basic statistics of the weekly network metrics in the new hire datasets.}
\end{table}

\begin{table}[h]
    \centering
    \small
\begin{subtable}[]{\linewidth}
\centering
\begin{tabular}{clll}
\toprule
\makecell[c]{Onboarding Week} &  Email &  IM  &  Meeting  \\
\midrule
0-6   &            0.147*** &                       0.138*** &                         0.035**                  \\
6-12  &            0.192*** &                      0.019 &                         0.005                  \\
12-18 &           -0.003 &                     0.087 &                         0.107               \\
18-24 &            0.320* &                       0.117 &                        0.080                \\

\bottomrule
\end{tabular}
   \caption{Differences of network structural diversity (*$p<0.05$, **$p<0.01$, **$p<0.001$).}
    \label{tab:topics_diff_ego_components_weighted}
\end{subtable}\\

\begin{subtable}[]{\linewidth}
    \centering
    \small
\begin{tabular}{clll}
\toprule
\makecell[c]{Onboarding Week} &  Email  &  IM &  Meeting \\
\midrule
0-6   &                 0.019*** &                             0.008 &                                 0.015***  \\
6-12  &                 0.013 *&                              0.021 ***&                                0.013***  \\
12-18 &                 0.024 ***&                             0.010 &                                0.002  \\
18-24 &                 0.002 &                                 0.006 &                                0.002  \\

\bottomrule
\end{tabular}
    \caption{Differences of the weekly growth of network structural diversity (*$p<0.05$, **$p<0.01$, **$p<0.001$).}
    \label{tab:topics_diff_ego_components_growth_weighted}
\end{subtable}
\caption{The relationship between new hires' engagements with the ``Pinned People'' and ``Suggested People'' feature in Microsoft Viva Topics and their network structural diversity  (\textbf{M4}), i.e., the number of ego components for group interaction (n<10).}

\label{tab:topics_diff_ego_components_overall_weighted}
\end{table}

\newpage

\begin{figure}[h]
    \centering
    \includegraphics[width=\linewidth]{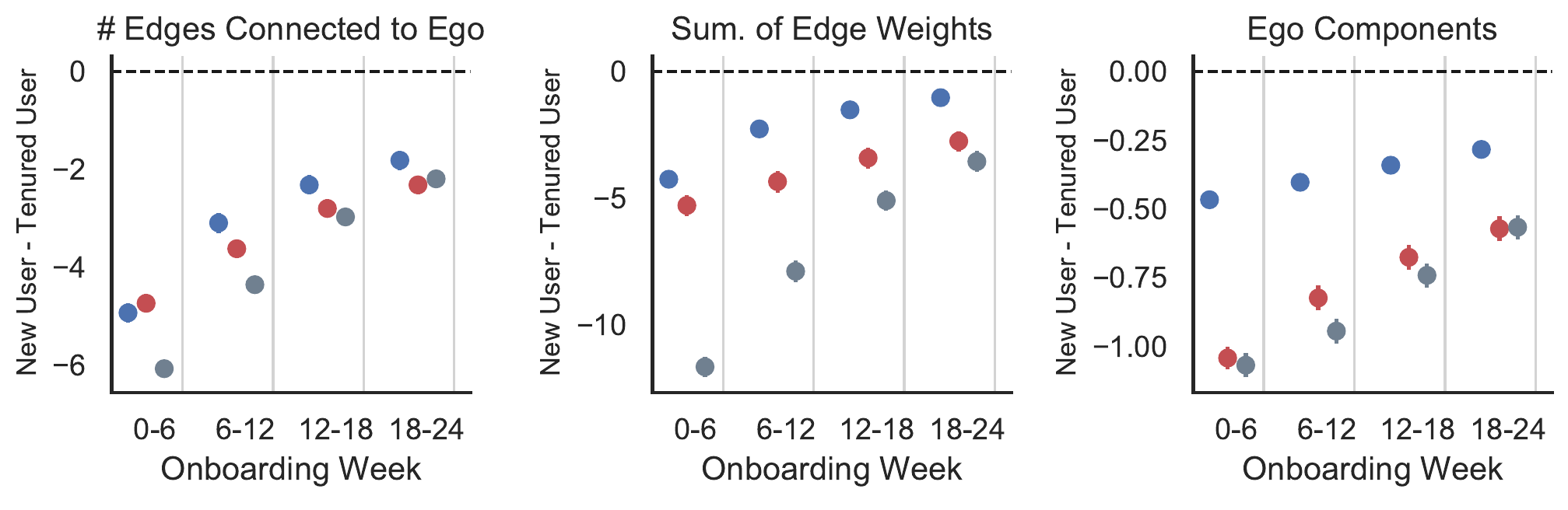}
    \caption{Differences of network statistics (with 95\% CIs) between new hires and tenured employees within each 6-week testing window (\textbf{M2}).}
     \Description[Differences of network statistics (with 95\% CIs) between new hires and tenured employees within each 6-week testing window (\textbf{M2}).]{Differences of network statistics (with 95\% CIs) between new hires and tenured employees within each 6-week testing window (\textbf{M2}).} 
    \label{fig:isNewhire_trends_weighted}
\end{figure}

\begin{figure}[h]
    \centering
\begin{subfigure}[b]{\linewidth}
    \centering
    \includegraphics[width=0.9\linewidth]{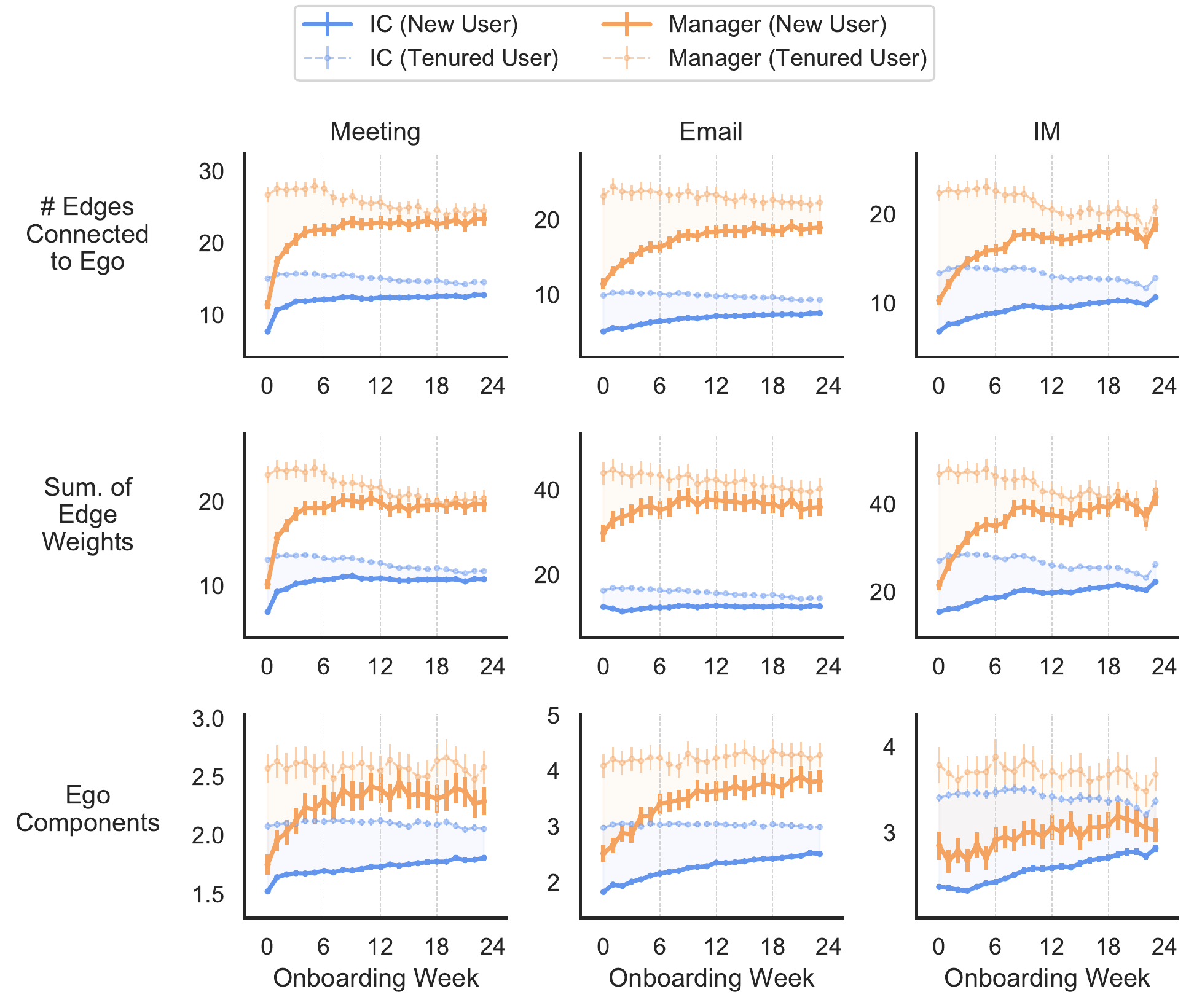}
    \caption{Heterogeneity w.r.t. Managerial Status (Manager vs. Individual Contributor).}
    
    \label{fig:isManager_trends_weighted}
\end{subfigure}

\begin{subfigure}[b]{\linewidth}
    \centering
    \includegraphics[width=0.9\linewidth]{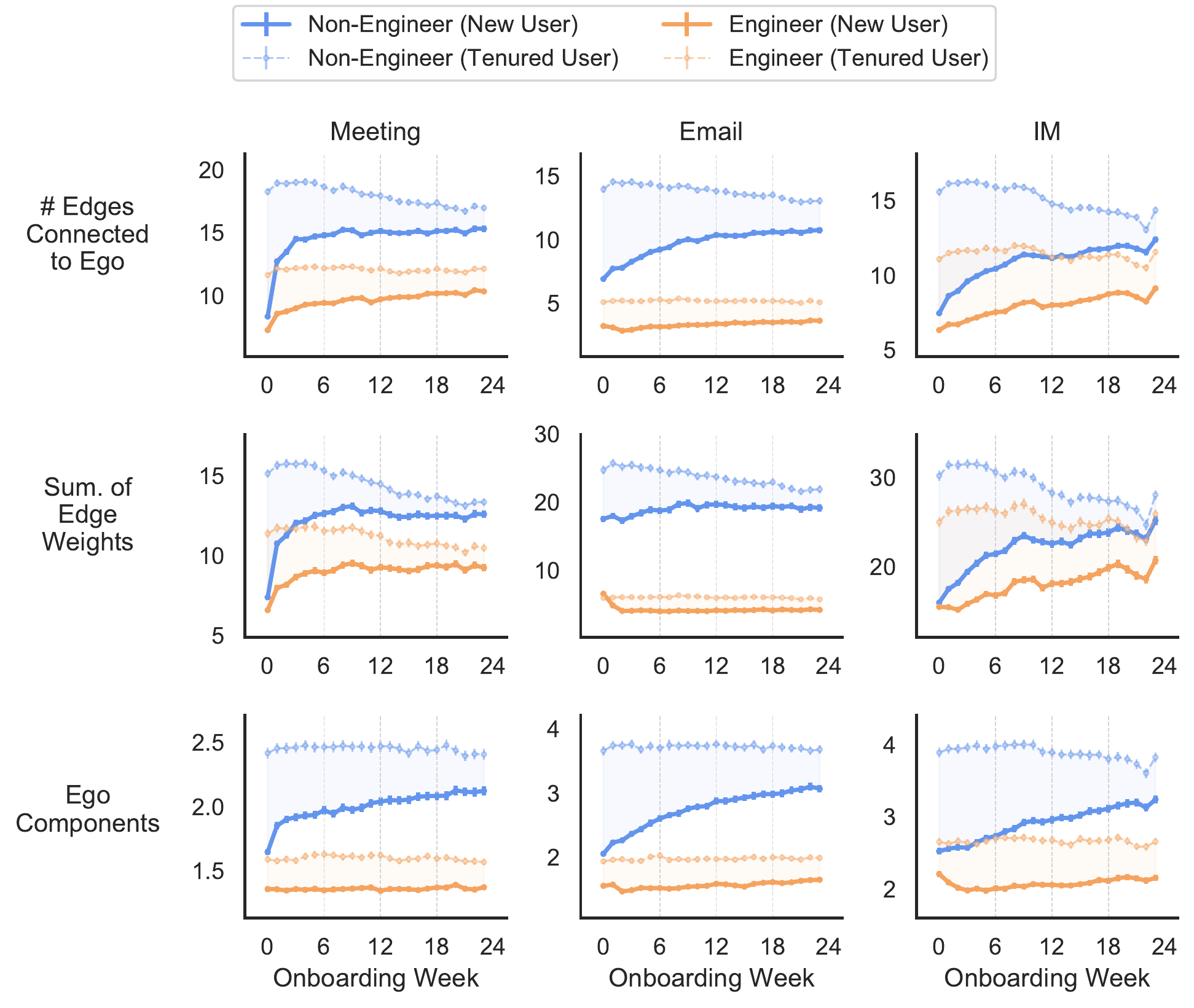}
    \caption{Heterogeneity w.r.t. Job function (Engineer vs. Non-Engineer).}
    \label{fig:isSWE_trends_weighted}
\end{subfigure}

\caption{Time trend comparisons across different new hire groups on their meeting, email and instant messaging networks of group interaction (n<10) (average network statistics with 95\% confidence intervals).}
\Description[Time trend comparisons across different new hire groups on their meeting, email and instant messaging networks of group interaction (n<10) (average network statistics with 95\% confidence intervals).]{Time trend comparisons across different new hire groups on their meeting, email and instant messaging networks of group interaction (n<10) (average network statistics with 95\% confidence intervals). 1) Heterogeneity w.r.t. Managerial Status (Manager vs. Individual Contributor).2)Heterogeneity w.r.t. Job function (Engineer vs. Non-Engineer).} 
    \label{fig:worker_heterogeneity_trends_weighted}
\end{figure}

\begin{figure}[h]
        \centering
\begin{subfigure}[b]{\linewidth}
    \centering
    \includegraphics[width=\linewidth]{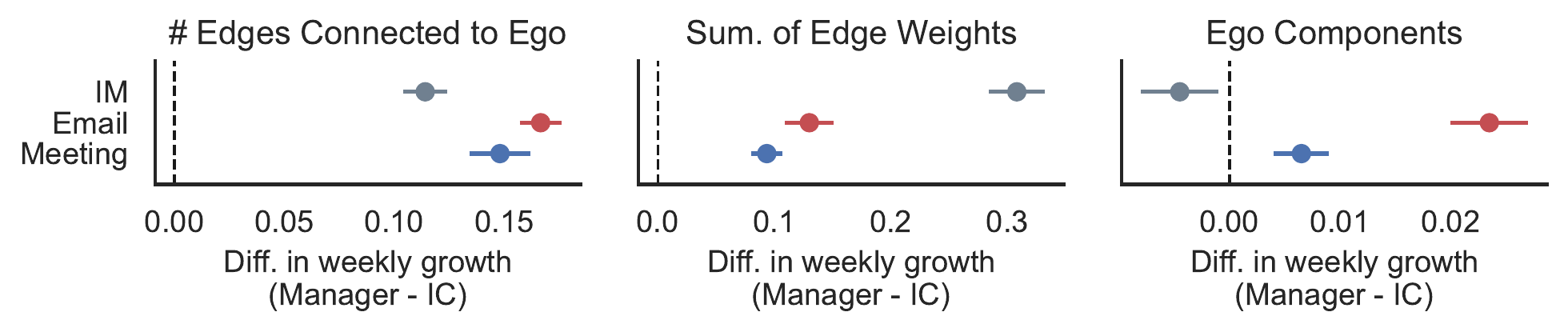}
    \caption{Manager vs. Individual Contributor.}\label{fig:CI_isManager_OnboardingWeek_overall_weighted}
\end{subfigure}

\begin{subfigure}[b]{\linewidth}
    \centering
    \includegraphics[width=\linewidth]{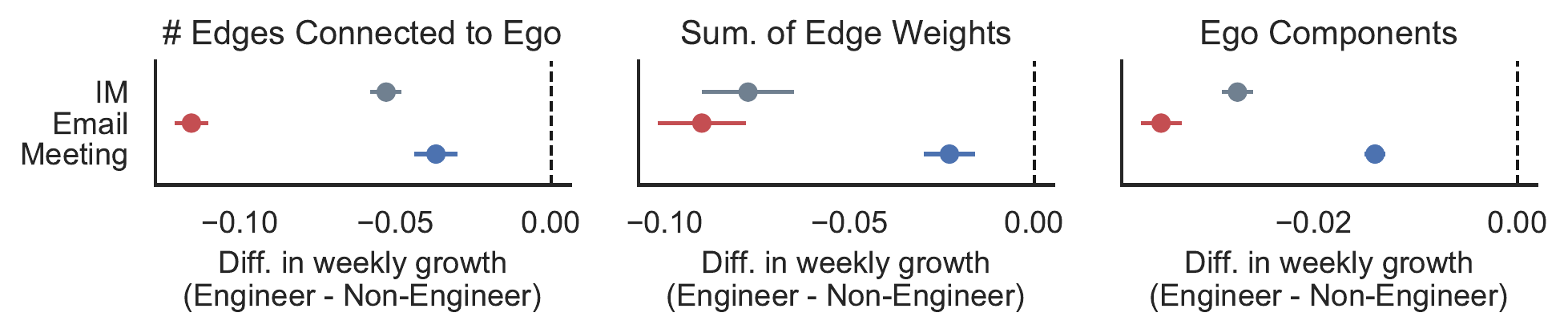}
    \caption{Engineer vs. Non-Engineer.}\label{fig:CI_isSWE_OnboardingWeek_overall_weighted}
\end{subfigure}

\caption{Heterogeneity of network growth rates (with 95\% CIs) of group interaction (n<10) across different new hire groups.}

 \Description[Heterogeneity of network growth rates (with 95\% CIs) of group interaction (n<10) across different new hire groups.]{Heterogeneity of network growth rates (with 95\% CIs) of group interaction (n<10) across different new hire groups. !)Manager vs. Individual Contributor. 2)Engineer vs. Non-Engineer.} 

\end{figure}

\begin{figure}[h]
    \centering
    \includegraphics[width=\linewidth]{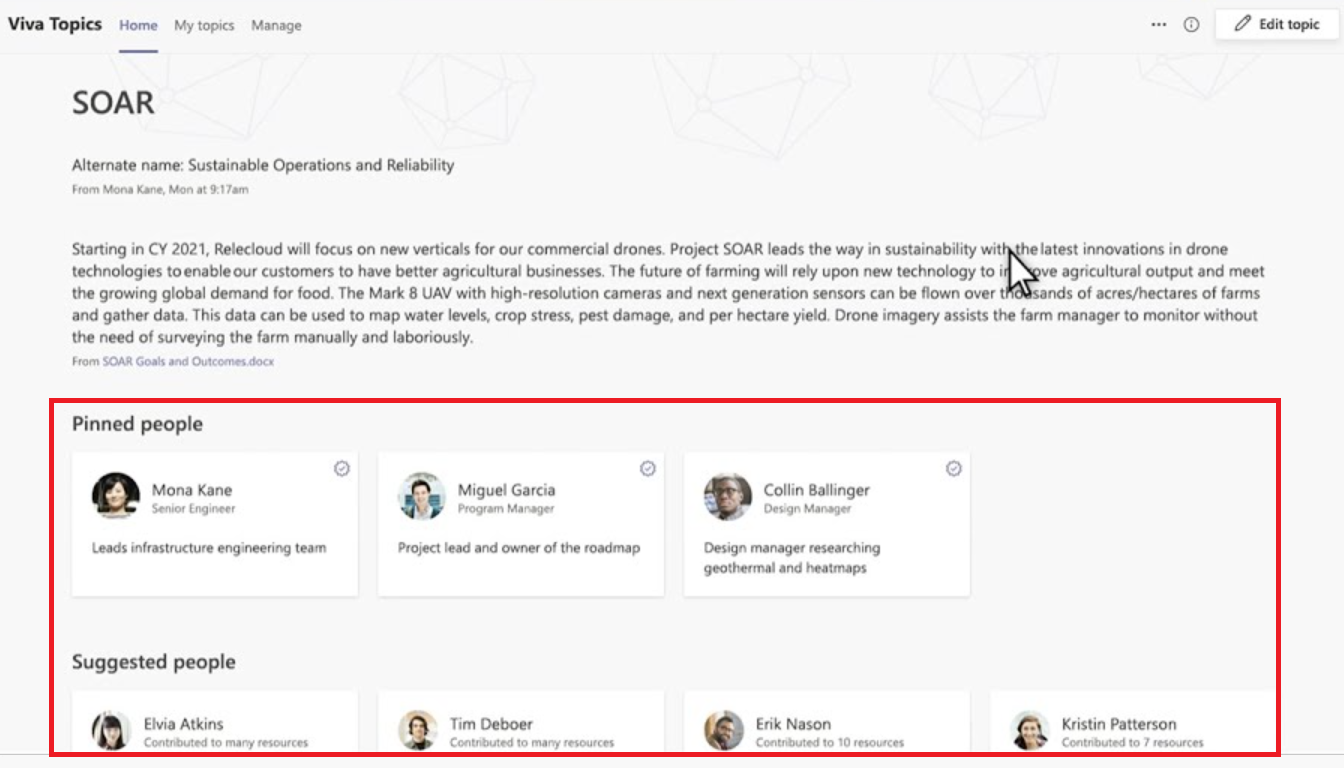}
    \caption{An illustration of the ``Pinned People'' and ``Suggested People'' features in Microsoft Viva Topics.}
    \Description[An illustration of the ``Pinned People'' and ``Suggested People'' features in Microsoft Viva Topics.]{An illustration of the ``Pinned People'' and ``Suggested People'' features in Microsoft Viva Topics.} 
    \label{fig:topics-mockup}
\end{figure}

\end{document}